\documentclass[twocolumn,prl,aps,flushbottom,showpacs]{revtex4-1}
\usepackage{xspace,amsmath,amsfonts,amsthm,amssymb,amsbsy,graphicx,color}

\begin{document}

\title{Ultra-Efficient Cooling of Resonators: Beating Sideband Cooling \\ with Quantum Control}

\author{Xiaoting Wang$^1$, Sai Vinjanampathy$^2$, Frederick W. Strauch$^3$, and Kurt Jacobs$^{2,4}$}

\affiliation{ 
$^1$Department of Applied Mathematics \& Theoretical Physics, University of Cambridge, Cambridge, CB3 0WA, UK \\ 
$^2$Department of Physics, University of Massachusetts at Boston, Boston, MA 02125, USA \\
$^3$Department of Physics, Williams College, Williamstown, MA 01267 \\
$^4$Hearne Institute for Theoretical Physics, Louisiana State
University, Baton Rouge, LA 70803, USA
}

\begin{abstract} 
The present state-of-the-art in cooling mechanical resonators is a version of ``sideband'' cooling. Here we present a method that uses the same configuration as sideband cooling --- coupling the resonator to be cooled to a second microwave (or optical) auxiliary resonator --- but will cool significantly colder. This is achieved by varying the strength of the coupling between the two resonators over a time on the order of the period of the mechanical resonator. As part of our analysis, we also obtain a method for fast, high-fidelity quantum information-transfer between resonators. 
\end{abstract}


\maketitle 

There is presently a great deal of interest in cooling high-frequency micro- and nano-mechanical oscillators to their ground states. This interest is due to the need to prepare resonators in states with high purity to exploit their quantum behavior in future technologies~\cite{Milburn08, Rabl10}.  The key measure of a cooling scheme is the \textit{cooling factor}, which we will denote by $f_{\mbox{\scriptsize cool}}$. The cooling factor is the ratio of the average number of phonons in the resonator at the ambient temperature, $n_T$, to the average number of phonons achieved by the cooling method, which we will denote by $\langle n \rangle_{\mbox{\scriptsize cool}}$. The present state-of-the-art for cooling mechanical resonators is \textit{sideband} cooling, which was originally developed in the context of cooling trapped ions~\cite{Wilson-Rae07, Marquardt07, Genes08}. This method is a powerful and practical technique, able to achieve large cooling factors, and these have been demonstrated in the laboratory~\cite{Gigan06, Thompson08, Groeblacher09, Park09, Lin09, Schliesser09, Rocheleau10, Teufel11, Riviere11, Li11}. 

In the context of mechanical resonators, sideband cooling involves coupling the resonator to be cooled (from now on the ``target'') to a microwave or optical resonator (the ``auxiliary'') whose frequency is sufficiently high that it sits in its ground state at the ambient temperature. The resonators are coupled together by a linear interaction, and one that is straightforward to implement experimentally. 
In particular, if we denote the annihilation operators for the target and auxiliary resonator by $a$ and $b$, respectively, then the full Hamiltonian of the two resonators is 
\begin{equation}
   H = \hbar \omega a^\dagger a + \hbar \Omega b^\dagger b + g \cos(\nu t) x_a x_b, 
   \label{eq::H}
\end{equation}
where $x_a = a + a^\dagger$ and $x_b = b + b^\dagger$ are the position operators of the respective resonators. The coupling is modulated at the difference frequency between the resonators, $\nu = \Omega - \omega$. This converts the high frequency of the auxiliary resonator so that the two resonators are effectively on-resonance, and thus exchange energy at the coupling rate $g$. With this frequency conversion, the auxiliary constitutes a source of essentially zero entropy (and thus zero temperature) for the target resonator~\cite{Tian09}. 

When the rate of the coupling, $g$, is significantly smaller than the frequency $\omega$ of the target resonator (so that one is within the rotating-wave approximation (RWA)--- see, e.g.~\cite{Irish07}), then the linear coupling between the resonators is merely excitation (phonon/photon) exchange between the two. If the auxiliary is now damped sufficiently rapidly, then the excitation exchange, combined with the relatively fast damping of the auxiliary at effectively zero temperature, extracts the phonons out of the target.  The cooling factor is merely the ratio of the phonon extraction rate to that of the heating rate, the latter being the product of the damping rate of the target, $\gamma$, multiplied by the average number of phonons it has at the ambient temperature, $n_T$. It is not always possible to determine the extraction rate analytically, but if we denote it by $\Gamma_{\mbox{\scriptsize cool}}$, then  
\begin{equation}
  \langle n \rangle_{\mbox{\scriptsize cool}} =  \frac{n_T}{f_{\mbox{\scriptsize cool}}}  = \left( \frac{\gamma}{\Gamma_{\mbox{\scriptsize cool}}} \right) n_T . 
\end{equation}
Note that $\Gamma_{\mbox{\scriptsize cool}}$ depends on how fast the auxiliary can extract energy from the target, or the ratio $g/\kappa$.  For sideband cooling, the RWA requires $g/\kappa \ll \omega/\kappa$, and thus limits the cooling factor.

Here we demonstrate that one can cool significantly better than traditional sideband cooling by using quantum control to go beyond the RWA, into the ultra-strong coupling regime $g \sim \omega$.  We first show that a particular time-dependence of the coupling rate, $g(t)$, can achieve a high-fidelity transfer of quantum states between the target and auxiliary resonators within a \textit{single} resonator period. As pointed out in~\cite{Jacobs11}, ``state-swapping'' is one way to achieve cooling, as this process will load the cold state of the auxiliary into the target. In fact, the phonon/photon exchange of the RWA implements state-swapping in a time of $\pi/(2g)$~\cite{Tian08}. However it was shown in~\cite{Jacobs11} that using this to cool (which means running traditional  sideband cooling, but now only for a single swap-time) is little better than the usual approach. In contrast, we show here that numerically optimized control sequences will achieve significantly better cooling factors.  Because this technique requires a relatively small modification to the existing sideband cooling scheme, it can be readily implemented with present technology. Further, our method not only provides better cooling, but makes this achievable over a much wider range of values of the auxiliary damping rate, $\kappa$. While our method always performs at least as well as sideband cooling for any value of $g$, to obtain the lowest achievable teperatures one does require ultra-strong coupling ($g\sim \omega)$. This is nevertheless timely, because very recent experiments have demonstrated coupling not far from this regime~\cite{Teufel11}. We note that Machnes~\textit{et al.}~\cite{Machnes10} have previously devised a way to go beyond the RWA in the context of trapped-ion experiments, in which the auxiliary system is a qubit. However, their method is not feasible for nano-resonators, certainly with present technology, because it requires $g\gg \omega$~\cite{Teufel11}.    


To begin our analysis we first consider the problem of engineering a fast, high-fidelity state-swap between two linearly coupled resonators, as this is an important problem in its own right. Fast operations on quantum information are important due to the the ever present effects of decoherence. To obtain such a state-swap, and thus an efficient energy transfer \textit{without} the RWA, we examine the algebra generated by the linear coupling in conjunction with the free Hamiltonians of the resonators.  The algebra of these three Hamiltonians suggests that it should be possible to engineer a perfect state-transfer operation between the two resonators, by concatenating the evolutions generated by the Hamiltonians in a process of ``quantum control''~\cite{DAlessandro08}. Up to local operations on each resonator, such a concatenation is equivalent to varying the coupling $g$ with time. This would allow us to obtain efficient energy transfer when $g \sim \omega$, not only achieving faster state-swapping, but also better cooling. 

To explore the above conjecture, we simulate the evolution given by the Hamiltonian in Eq.(\ref{eq::H}), in which $g$ is a function of time. Since $\Omega$ is typically much greater than $\omega$ (by a factor of at least 100), it is a good approximation to assume that the frequency conversion is exact, and set $\Omega = \omega$ and $\nu = 0$. The corrections to this approximation are of the order of $(\omega/\Omega)^2$. (This is, in fact, an RWA for the frequency $\Omega$, which is distinct from the RWA for the target frequency $\omega$, required by sideband cooling.) We prepare the target resonator in a state that is confined to the space spanned by the $12$ lowest Fock states, but otherwise completely mixed on that space. The auxiliary is prepared in the ground state, and the resonators evolved for a specified time. This allows us to determine the quality of the swap merely by calculating the purity of the final density matrix for the target resonator. If this state is pure, then the evolution has successfully transferred all the quantum information to the auxiliary resonator. We evolve for a single period of the target resonator, and dividing this time into five equal intervals of duration $\Delta t$, we parametrize $g(t)$ by making it piecewise-constant on these intervals. Finally we perform a numerical optimization, using a Quasi-Newton line search method~\cite{Nocedal06}, to determine the five piecewise-constant values for $g(t)$. For the simulation we use the basis of Fock states, including the lowest 25 states for each resonator. This achieves an essentially perfect state-swap (a final purity of $0.999977$) with the following five values of $g/\omega$: $(1.78, 1.45, 2.44, 1.61, 0.195)$. As a second example, we find that a state-swap with a purity of $0.999991$ can be obtained in $0.7$ of the resonators period, with the values $(2.76,0.474,3.73,0.78,2.59)$. 

The above results show that, in the absence of decoherence, state-swapping in less than one period is within the ``control space'' of the linear coupling. But this does not tell us how well we can transfer the cold auxiliary state to the hot resonator in the presence of damping. Damping interferes with the swapping process via the quantum Zeno. Damping is equivalent to a continuous measurement process~\cite{Jacobs11,JacobsSteck06}, and this inhibits the transfer of energy to the auxiliary. We must therefore simulate the optimized cooling in the presence of damping, but it is impractical to do this with the simulation method used above, as the size of the required superoperators is too large. Fortunately in the case of cooling we are only interested in the average phonon number, given by $\langle n \rangle = \langle a^\dagger a \rangle$, which is a second moment of operators $a$ and $a^\dagger$. Because the dynamics of the resonators is linear (that is, the evolution can be described by a set of linear quantum Langevin equations~\cite{Clerk08,Clerk10,QNoise}) one can derive a closed set of equations for the variances and covariances of the annihilation operators. Because the means of these operators are zero in thermal states, and remain zero during the evolution, the covariances are equal to the second moments. 

\begin{figure*}
\leavevmode\includegraphics[width=1\hsize]{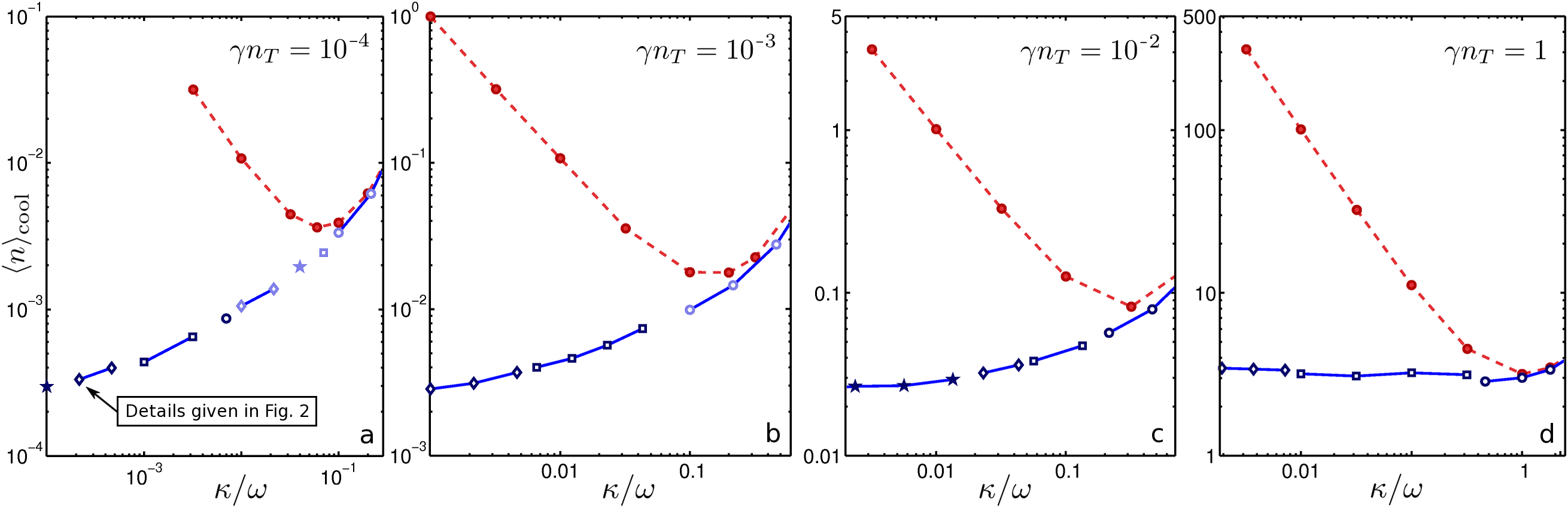}
\caption{(Color online) The average phonon number, $\langle n \rangle_{\mbox{\scriptsize cool}}$, achieved by the cooling method presented here, compared to sideband cooling, as a function of the damping rate of the auxiliary resonator, $\kappa$, and for four values of $\gamma n_T$. The values obtained by sideband cooling, for the optimal choice of coupling $g$, are given by the solid circles/dashed line (red online). The lower curves, consisting of various symbols, give the values achieved by the present cooling method, in which $g$ is varied with time. The various symbols indicate different values of the total time, $\tau$, over which the cooling is achieved. In the following, $\tau$ is in units of oscillator period. (a) From left to right: Star: $\tau= 0.5$; Diamonds: $\tau= 0.6$; Squares: $\tau=1$; Circle: $\tau=1.5$; Diamond: $\tau= 2$; Star: $\tau= 3$; Squares: $\tau= 5.3$; Circles: $\tau=6$. (b) Diamonds: $\tau= 0.7$; Squares: $\tau= 1$; Circles: $\tau=6$; (c) Stars: $\tau= 0.6$; Diamonds: $\tau= 0.8$; Squares: $\tau= 1$; Circles: $\tau= 2$; (d) Diamonds: $\tau= 1$; Squares: $\tau= 2$; Circles: $\tau= 6$.} 
\label{fig1} 
\end{figure*} 

If we define the vector $\mathbf{x} \equiv (a, a^\dagger, b, b^\dagger)^{\mbox{\scriptsize t}}$, then the matrix of covariances is $C \equiv \langle \mathbf{x}\mathbf{x}^{\mbox{\scriptsize t}} \rangle - \langle \mathbf{x} \rangle \langle \mathbf{x}^{\mbox{\scriptsize t}} \rangle$. The equation of motion for $C$ is 
 \begin{equation}
    \dot{C} = A C + C A^{\mbox{\scriptsize t}} + G  , 
    \label{eq::mom2}
\end{equation}
where 
\begin{eqnarray}
    A & = &  \left( \begin{array}{cccc}  -i\omega - \frac{\gamma}{2} & 0 & - i g & - i g  \\  0 &  i\omega - \frac{\gamma}{2} & i g &  i g  \\ -ig & -ig & -i\omega - \frac{\kappa}{2} & 0 \\ ig & ig & 0 & i\omega - \frac{\kappa}{2} \end{array} \right) , \\ 
    G & = &  \left( \begin{array}{cccc} 0 & \gamma (n_T+1) & 0 & 0  \\  \gamma n_T & 0 & 0 & 0  \\ 0 & 0 & 0 & \kappa (n_{\mbox{\scriptsize aux}} + 1) \\ 0 & 0 & \kappa n_{\mbox{\scriptsize aux}} & 0 \end{array} \right) , 
\end{eqnarray}
and $n_{\mbox{\scriptsize aux}}$ is the initial average number of phonons in the auxiliary resonator.  

We now use the above equations to perform a numerical optimization as before, but this time to determine the function $g(t)$ that gives the minimum value of $\langle a^\dagger a \rangle$ after a fixed time interval. To do this we take the same approach as above, dividing the total evolution time into $N$ equal intervals, and making $g$ piecewise constant on those intervals. We wish to determine the optimal cooling over a broad range of the relevant parameters, and compare this to sideband cooling. The important parameters are the damping rates of the target and auxiliary (respectively $\gamma$ and $\kappa$), and the average number of thermal photons in the target at temperature $T$, $n_T$. We note that the average number of photons in the auxiliary the ambient temperature, $n_{\mbox{\scriptsize aux}}$, can be made very small with present technology. For example, a $10 \,\mbox{GHz}$ stripline resonator at $50 \,\mbox{mK}$ has $n_{\mbox{\scriptsize aux}} = 6.7\times 10^{-5}$. The same resonator at $100 \,\mbox{mK}$ has $n_{\mbox{\scriptsize aux}} = 8.3\times 10^{-3}$. We expect $n_{\mbox{\scriptsize aux}}$ only to be significant if we are able to cool the target so that $\langle a^\dagger a\rangle$ is close to $n_{\mbox{\scriptsize aux}}$, and we verify this below. 
 
By examining the equations of motion, we see that so long as $n_T \gg 1$, the evolution, and thus the cooling, depends only on the product of $\gamma$ and $n_T$, rather than each separately. Since $n_T \gg 1$ is the relevant regime for present experiments, we therefore need to determine the optimal cooling as a function of $\kappa$ and $\gamma n_T$. Finally, it turns out that there is not only an optimal $g(t)$ for each value of $\kappa/\omega$, but also an optimal \textit{time} over which to perform the control. We could optimize $g(t)$ and the total time in a single shot, but we found that this extended the optimization time considerably. We therefore perform the optimization for a set of fixed total times, and plot the best results for each value of $\kappa$. 
 
We now perform the optimization over $g(t)$, with $n_{\mbox{\scriptsize aux}} = 0$, and plot the results in Figure~\ref{fig1}. It is convenient to refer to $\gamma$ and $\kappa$ in units of $\omega$, and thus we plot the achieved value of $\langle a^\dagger a\rangle$ as a function of $\kappa/\omega$, and for four values of the product $(\gamma/\omega) n_T$. In Figure~\ref{fig1} we also plot, for direct comparison, the values of $\langle a^\dagger a\rangle$ that are achieved using sideband cooling. In agreement with~\cite{Genes08}, we find that sideband cooling achieves its best performance when $\kappa$ is in the range $0.1\omega$ -- $\omega$.  

We see from Fig.~\ref{fig1} that our ``optimal control'' cooling scheme is superior to sideband cooling when $\kappa$ is \textit{less than} the value for which sideband cooling achieves its best performance; above that value of $\kappa$ their performance is very similar. The second key result is that the improvement provided by optimal control steadily increases as the product $\gamma n_T$ decreases with respect to $\omega$. For $(\gamma/\omega) n_T = 10^{-4}$, $10^{-3}$, and $10^{-2}$, the smallest values we obtained for $\langle a^\dagger a\rangle$ are better than sideband cooling by factors of approximately $12$, $6$, and $3$, respectively (see Figs.~\ref{fig1}(a)-(c)). For $(\gamma/\omega) n_T = 1$, while our method no longer achieves a lower temperature than sideband cooling, it achieves this temperature for a wide range of values of $\kappa$, whereas sideband cooling does so only for a single value.    

\begin{figure}
\leavevmode\includegraphics[width=1\hsize]{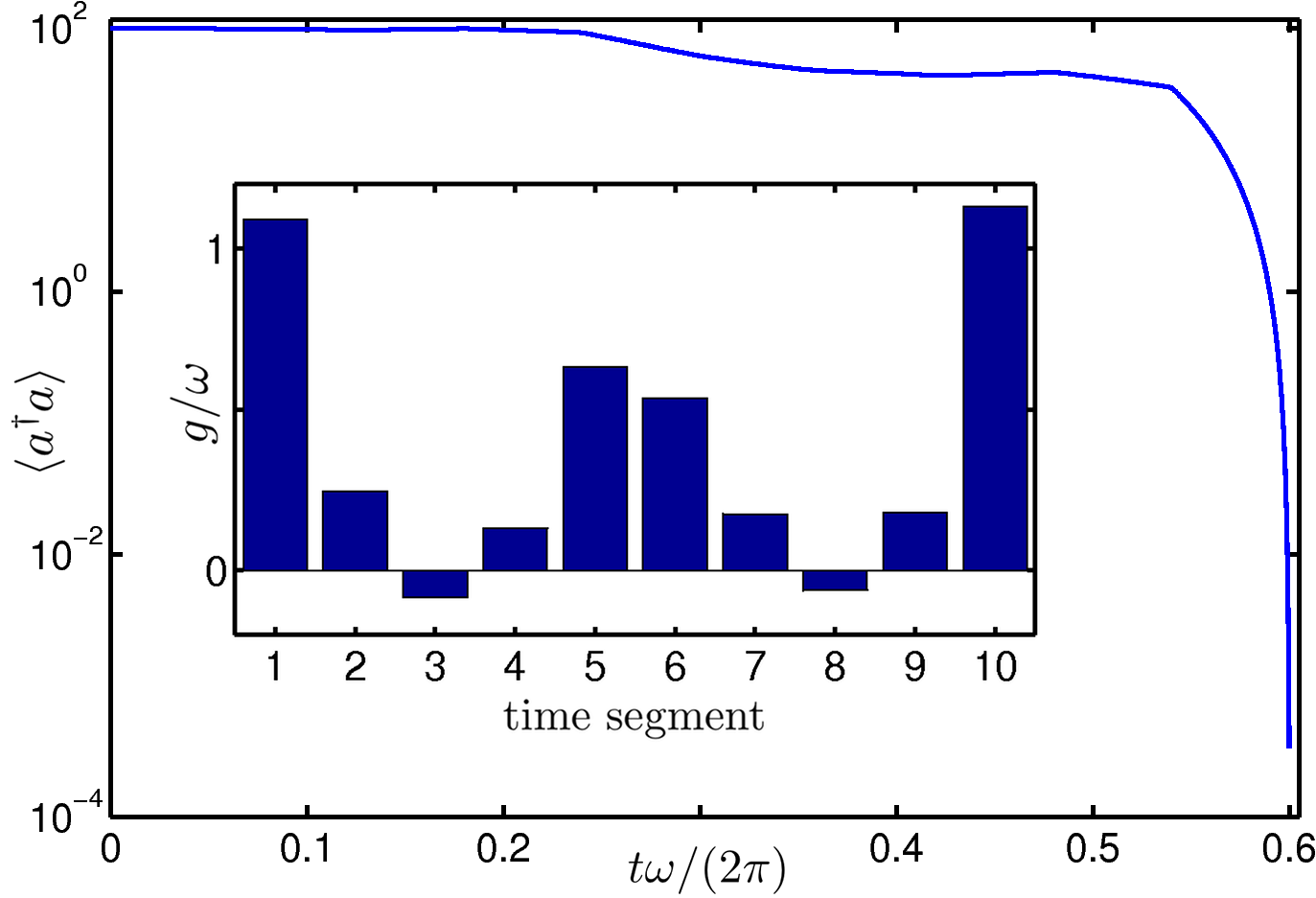}
\caption{(Color online) Here we show the optimal 10-segment pulse shape for the interaction rate $g(t)$ (inset), along with the time-dependence of the average number of photons, for one of the points in Fig.~\ref{fig1}b. In this case $n=100$, $\gamma = 1e-6\omega$, $\kappa = 1.35\times 10^{-3}\omega$, and the duration of of the control pulse is $0.6$ periods of the resonator.} 
\label{fig2} 
\end{figure}

Almost all the cooling results in Fig.~\ref{fig1} are obtained with no more than 20 time-segments (that is, 20 piecewise-constant values for $g(t)$) per period. The exception is the diamonds in Fig.~\ref{fig1}(b), in which we used 30 segments in $0.7$ of a period. In many case we find that between 5 and 10 time-segments per period is sufficient to obtain optimal cooling. In Fig.~\ref{fig2} we show the explicit control for $g(t)$, for a chosen value of $\kappa$ in Fig.~\ref{fig1}a, along with the evolution of $\langle a^\dagger a\rangle$ for this sequence. While we have chosen to parametrize $g(t)$ using a piecewise constant function, one can just as easily use a Fourier series, which removes the need for sharp switching between the segments. When optimality is achieved with 20 segments per period, this indicates that the highest frequency component required for $g$ in such a Fourier series is no more than $20\omega$. For a $50\, \mbox{MHz}$ resonator, this is $1\,\mbox{GHz}$, and thus experimentally quite practical.  We note that as $\kappa$ increases, the time, $\tau$, required for optimal cooling also increases. For $\kappa = \omega/1000$, $\tau \sim 0.7$ of the resonator period, whereas when $\kappa = \omega/2$, $\tau$ is typically about six periods (the specific cooling times are given in the caption of Fig.~\ref{fig1}).   

We now determine the effect of thermal photons in the auxiliary resonator ($n_{\mbox{\scriptsize aux}} > 0$). We perform the optimization again for a case in which we obtain the lowest temperatures: $\gamma = 10^{-6}\omega$, $n_T = 100$, and five values of $\kappa$ equally spaced on a log scale from $10^{-4}\omega$ to $10^{-3}\omega$. First we set $n_{\mbox{\scriptsize aux}} = 0$ and obtain the following set of cooled values for $\langle a^\dagger a \rangle$: $10^{-4}\times(2.9,4.0,4.3,5.3,7.0)$. Now performing the optimization with $n_{\mbox{\scriptsize aux}} = 10^{-4}$, the new set of cooled values for $\langle a^\dagger a \rangle$ are $10^{-4}\times(4.1, 4.4, 5.0, 6.1, 8.4)$. The increase in the average phonon number is approximately the addition of $n_{\mbox{\scriptsize aux}}$. This confirms our intuition that the effect of thermal photons in the auxiliary is only significant when the target is cooled close to $n_{\mbox{\scriptsize aux}}$. 

Finally, we consider coupling two auxiliary oscillators to the target resonator, in which both coupling rates are independent functions of time. Performing an optimization in this case, we find that this configuration does not provide any significant improvement in the achievable cooling. This suggests that the cooling method we have presented gives the best cooling that is possible using linear elements. 

{\em Acknowledgements:} FWS and KJ are supported by the NSF under Project No. PHY-1005571. SV and KJ are supported by the NSF under Project No. PHY-0902906, and KJ is also supported by The Hearne Institute for Theoretical Physics, ARO and IARPA. 


%

\end{document}